\documentclass[sigconf]{acmart}
\usepackage[utf8]{inputenc}
\usepackage{url}
\usepackage{multirow}
\usepackage{nicefrac}
\usepackage{mathell}
\usepackage{tilde}% modifies tilde in mathmode, use ~sin like \sin with any letter sequence
\usepackage{algorithm}
\usepackage{algpseudocode}
\usepackage{subcaption}
\usepackage{graphicx}

\let\xxx\epsilon\let\epsilon\varepsilon\let\varepsilon\xxx
\let\xxx\phi\let\phi\varphi\let\varphi\xxx

\def\hat#1{#1'}
\let\Tilde\widetilde
\newcommand{\mpara}[1]{\par\noindent{\bf #1\@.}}

\title{W-Trace: Robust and Effective Watermarking for GPS Trajectories}
\setcopyright{none}
\author[Rajjat Dadwal, Thorben Funke, Michael Nüsken, Elena Demidova]{Rajjat Dadwal$^{1}$, Thorben Funke$^{1}$, Michael Nüsken$^{2}$, Elena Demidova$^{3}$}
\affiliation{%
  \institution{
  $^{1}$L3S Research Center, Leibniz University Hannover, Hannover\country{Germany} \\
  $^{2}$Bonn-Aachen International Center for Information Technology, Bonn\country{Germany}\\
  $^{3}$Data Science and Intelligent Systems Group (DSIS), University of Bonn, Bonn\country{Germany}
  }
  }
\email{dadwal@L3S.de, tfunke@L3S.de, nuesken@bit.uni-bonn.de, elena.demidova@cs.uni-bonn.de}

\newcommand\blfootnote[1]{%
  \begingroup
  \renewcommand\thefootnote{}\footnote{#1}%
  \addtocounter{footnote}{-1}%
  \endgroup
}

\newcommand{\approach}{\textit{W-Trace}}

\renewcommand\footnotetextcopyrightpermission[1]{}
\AtBeginDocument{
  \providecommand\BibTeX{{%
    \normalfont B\kern-0.5em{\scshape i\kern-0.25em b}\kern-0.8em\TeX}}}

\setcopyright{None}

\begin{document}
\begin{abstract}
With the rise of data-driven methods for traffic forecasting, accident prediction, and profiling driving behavior, personal GPS trajectory data has become an essential asset for businesses and emerging data markets. However, as personal data, GPS trajectories require protection. 
Especially by data breaches, verification of GPS data ownership is a challenging problem. 
Watermarking facilitates data ownership verification by encoding provenance information into the data. 
GPS trajectory watermarking is particularly challenging due to the spatio-temporal data properties and easiness of data modification; as a result, existing methods embed only minimal provenance information and lack robustness. 
In this paper, we propose \approach{} -- a novel GPS trajectory watermarking method based on Fourier transformation. 
We demonstrate the effectiveness and robustness of \approach{} 
on two real-world GPS trajectory datasets.

\end{abstract}


\maketitle
\pagestyle{plain}
\blfootnote{\textcopyright Rajjat Dadwal, Thorben Funke, Michael Nüsken and Elena Demidova, 2022. This is the author's version of the work. It is posted here for your personal use. Not for redistribution. The definitive version was published in the proceedings of 
The 30th International Conference on Advances in Geographic Information Systems, SIGSPATIAL 2022, \url{https://doi.org/10.1145/3557915.3561474}.\\}
\section{Introduction}
\label{sec:introduction}
Personal GPS trajectory data are adopted in various critical domains, including data-driven urban traffic management, mobility, communication, and health. 
However, GPS trajectory data encode sensitive personal information such as user addresses, visited locations, and routes. 
Sharing and trading personal GPS trajectory data, even based on user consent, can occasionally result in data breaches and user privacy loss \citep{chow2011trajectory}. 
\begin{figure}[h!]
   \centering
     \includegraphics[width=1\linewidth,height=4.6cm]{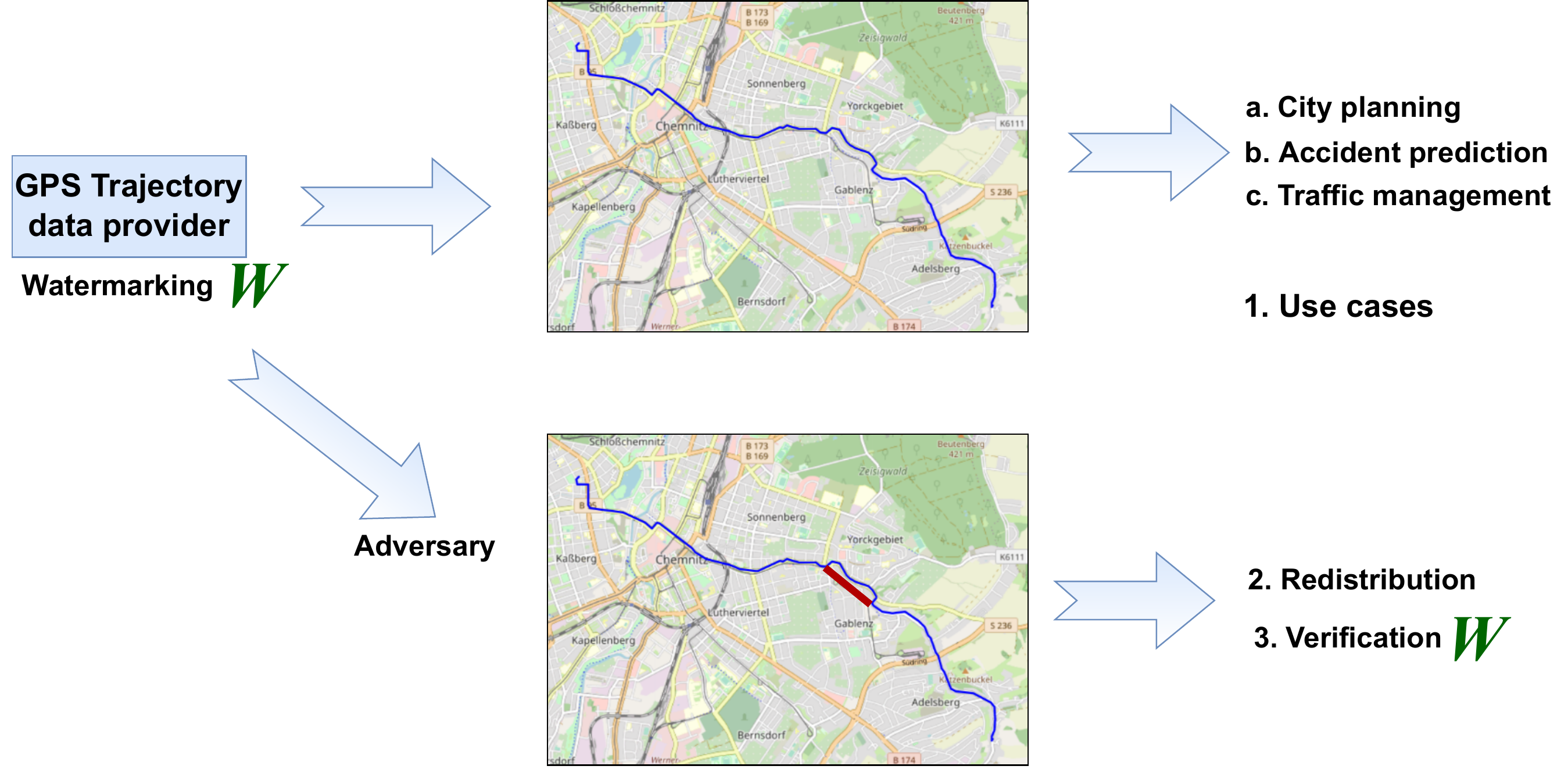}
    \caption{An example \approach{} application scenario. Watermarked GPS trajectory data is modified and re-distributed by an adversary. 
    \approach{} enables data provenance verification.
    Map data: \textcopyright OpenStreetMap contributors, ODbL.
    }
    \label{fig:approach3k}
\end{figure}

Figure \ref{fig:approach3k} illustrates an example application scenario in which GPS trajectory data, initially shared according to the user's consent, is obtained by an adversary due to a data breach, modified to obscure the data origin, and illegally re-distributed on the market.
Whereas the modification makes it challenging to claim the data ownership and to identify the misuse, sensitive personal information, such as user routes and driving patterns encoded in the trajectory, remains visible. 
The risk of data breaches necessitates the development of effective and robust provenance information embedding methods for personal GPS trajectory data to facilitate data provenance verification. 

Digital watermarking refers to methods that embed provenance information (so-called watermarks) into noise-tolerant data.
Watermarking has been extensively studied in the media domain to protect images, videos, and audio files~\citep{bhat2011new,DBLP:journals/ijst/El-WahabEAE21}. 
In contrast, only a few initial approaches target watermarking of personal GPS trajectories~\citep{pan2019trajguard,jin2005watermarking}.
Watermarking GPS trajectories poses several challenges and is an inherently difficult task. 
The strength of a watermark is subject to a trade-off. 
On the one hand, a watermark should be robust, i.e., strong enough not to be removed by an adversary. 
On the other hand, a watermark should, at the same time, be weak, such that the watermarked data is still usable in the downstream applications.
In addition to this general challenge for digital watermarking, GPS trajectories are, with their non-uniform sampling rate and positional inaccuracy, inherently susceptible to different modifications than media data, such as removal/addition of points or re-sampling along the path.  
State-of-the-art watermarking methods in the trajectory domain either lack robustness \citep{jin2005watermarking} or are ineffective, i.e., they embed only a small amount of data \citep{pan2019trajguard}. 

In this paper, we propose \approach{} -- a novel, robust and effective watermarking method for personal GPS trajectories.
\approach{} represents two-dimensional trajectory coordinates as complex numbers and adopts Discrete Fourier Transform (DFT) to enable effective watermark embedding in the frequency domain. To the best of our knowledge, we are the first to propose a DFT-based watermarking scheme for GPS trajectories.
We confirm the effectiveness and robustness of our approach by considering a comprehensive set of attacks, i.e., adversarial trajectory modifications, including noise addition, point replacement, and size modifications.
We conduct an extensive evaluation using two real-world GPS trajectory datasets.
We demonstrate that under the majority of considered attacks, \approach{} retains the watermark in 100\% cases.
We make our algorithm and data processing pipeline available as open source\footnote{Software: \url{https://github.com/Rajjat/watermarkingTrajectory}}.  

\section{Definitions \&{} Problem Formulation}
\label{sec:preliminaries}
In this section, we introduce the definitions and the problem formulation, which we tackle with the proposed \approach{} approach. 
\begin{definition}[Trajectory]
    A trajectory $T$ is a list of GPS coordinates ordered by the corresponding timestamps:
  \[
       T=[ (p_j,t_j)], \mbox{with $t_j$ < $t_{j+1}$ for all $j$},
      \]
    where $p_j = (a_j, b_j)$ is the two-dimensional position with latitude $a_j$ and longitude $b_j$ and $t_j$ is the timestamp of that position.
   Trajectory size, $\operatorname{size}(T)$, denotes the number of timestamps included in $T$. 
\end{definition}
A watermark is a signal embedded into the trajectory to enable verification of the trajectory origin. 
In this work, we represent watermarks as integer vectors.
\begin{definition}[Watermark]
A watermark $w \in \mathbb{Z}^{m}$ is an integer vector with the dimensionality $m$.
\end{definition}
The dimensionality $m$ of the watermark corresponds to the size of the (sub-)trajectory in which the watermark is embedded.

\textbf{Watermark verification} confirms if a given original watermark is embedded into the data and requires both the extracted watermark and the original watermark to be verified.
 
When the watermarking process  modifies a trajectory $T$ into $\Tilde{T}$,  
$\Tilde{T}$ needs to maintain usability for real-world applications. 
We make that intuition precise by defining a modification threshold.

\begin{definition}[Modification threshold]
    \sloppy%
    A modification threshold~$\sigma$ bounds a distance $D$ for trajectories.
    Given a modification threshold~$\sigma$, we consider $\Tilde{T}$ a $\sigma$-modification of $T$ if the spatial distance between these two trajectories is at most~$\sigma$.
    Formally:
   \begin{equation} 
       \operatorname{D}(T,\Tilde{T}) \leq \sigma.
       \label{eq:mod-threshold}
    \end{equation}
\end{definition}
In our experiments, we work with $\sigma=$10~meters, which reflects the typical inaccuracy of GPS sensors~\citep{bevly2004global}.

Our goal is to watermark GPS trajectories such that the watermarked trajectory remains usable for downstream applications and the watermark can be verified effectively, even if the watermarked trajectory is modified. 
Formally, given a watermark embedding procedure $~EMB$, the respective watermarking verification procedure $~VER$, and a watermark $w$, 
we aim that a trajectory $T$ and its corresponding watermarked trajectory $\Tilde{T}=~EMB(T, w)$ obtained after applying watermarking are within the predefined modification threshold~$\sigma$. 
Moreover, we aim that the verification of $w$ with $~VER$ is possible, even if $\hat{\Tilde T}$ is modified from $\Tilde{T}$ within a modification threshold~$\sigma$.
Hence, we want to ensure that the verification $~VER(\hat{\Tilde T}, T, W)$ returns true, if $\hat{\Tilde T}$ is a $\sigma$-modification of $\Tilde{T}$. 

\section{The \approach{} Approach}
\label{sec:approach}

This section presents our proposed watermark embedding and verification method \approach{}. 

\subsection{Watermark Embedding}
\label{sec:watermark addition}
Watermark embedding aims to incorporate a watermark into a given GPS trajectory. 
We consider a trajectory $T$ of size $n$. 
We associate each GPS point $(a_j, b_j)$ with a complex number, 
\begin{align}
    c_{j} = a_{j} + i b_{j}, \label{eq:complex}
\end{align}
where $i$ is the imaginary unit.
We split the transformed trajectory into multiple sub-trajectories of equal size.
Next, we apply a Discrete Fourier Transform (DFT) to each sub-trajectory, where we use the Fast Fourier Transform (FFT) \citep{nussbaumer1981fast} algorithm for efficiency.
DFT retrieves a frequency domain representation of the input and results in a sequence of complex numbers of the same length as the input. 
We feed the list of positions $c = (c_{j})_{k\le j<l}$ from the sub-trajectory spanning the indices $k$ to $l$, represented as complex numbers, into the FFT algorithm. 
The resulting frequency representations we then represent via amplitudes $\alpha$ and phase angles $\phi$:
\begin{equation} 
    \label{eq:FFT}
    \alpha,\phi \gets ~FFT(c).
\end{equation}
Then, for a sub-trajectory, the watermark~$w$ with strength $s$ is inserted in the amplitude~$\alpha$:
\begin{equation} 
    \label{eq:watermark}
    \Tilde{\alpha} = \alpha+s \cdot w.
\end{equation}
A design decision of our method is to represent the watermark~$w$ as a vector of $1$, $-1$, and $0$ values
of the same size as each sub-trajectory. 
This watermark is chosen and stored by the user; the watermark may be the same for each sub-trajectory or vary. In our experiments, we generate the watermarks randomly.
The higher the watermark strength $s$, the more we modify the trajectory by inserting the watermark.
In our experiments, we use $s=0.0003$.
We split each trajectory into sub-trajectories of size 16. 
     In each sub-trajectory, we embed a watermark with $10$ non-zero dimensions.
Once the watermarks are inserted in the amplitude of each sub-trajectory, the next step is to apply an inverse FFT (IFFT) to obtain the watermarked sub-trajectory.
We take the watermarked amplitude $\Tilde{\alpha}$ with the original phase~$\phi$ and form a complex number $t_{j} \gets \Tilde{\alpha}_{j} ~exp( i \phi_{j} )$.
Applying the inverse FFT to the vector $t$, we obtain the watermarked trajectory
~$\Tilde{c}$. We abbreviate this as follows:
\begin{equation} 
    \label{eq:watermarkinserted}
   \Tilde{c}=(\Tilde{a},\Tilde{b}) \gets ~IFFT( \Tilde{\alpha}, \phi ).
\end{equation}

\subsection{Watermark Extraction \& Verification}
\label{sec:watermark-extraction}

Watermark verification aims to verify if the specific watermark $w$ is embedded in the given trajectory $\hat{\Tilde T}$. This process includes four steps: selection of a candidate trajectory, trajectory size alignment, watermark extraction, and watermark correlation.

\textbf{Candidate selection.} As input, the watermark verification process requires 
the trajectory $\hat{\Tilde T}$ to be verified, 
the original trajectory $T$, 
the watermark $w$ and the watermark strength parameter $s$ 
adopted in the watermark embedding process. 
As the candidate original trajectory $T$, we select the closest user trajectory based on the minimum haversine distance to $\hat{\Tilde T}$.

\textbf{Trajectory size alignment.} Our watermark verification process requires $T$ and $\hat{\Tilde T}$ to be of the same size.
If $size(\hat{\Tilde T})>size(T)$, i.e. the trajectory size increased, we filter the coordinates from $\hat{\Tilde T}$ based on the minimum haversine distance to the candidate trajectory $T$.
If the trajectory size of $\hat{\Tilde T}$ is smaller than $size(T)$, we fill the positions in $\hat{\Tilde T}$ with a re-sampling of the closest point (regarding the haversine distance) to obtain the same size.

\textbf{Watermark extraction}. 
The watermark extraction process in \approach{} is non-blind, 
i.e., requiring the original data,
and is the reverse of the watermark insertion process.
We split $\hat{\Tilde T}$ into sub-trajectories of equal size and apply DFT to calculate the amplitude $\hat{\alpha}$.
We retrieve the watermark with:

\begin{equation} 
    \label{eq:watermark1}
    w'=\frac{\hat{\alpha}-\alpha}{s},
\end{equation}
where $\alpha$ is the amplitude of the candidate trajectory $T$ and $s$ is the watermark strength.

\textbf{Watermark correlation.} The next step to verify the watermark is to compute the correlation between the extracted watermark $\hat{w}$ and the original watermark $w$ of each sub-trajectory.
We adopt Normalized Cross-Correlation (NCC) -- a widely used watermark verification measure \citep{DBLP:journals/ijst/El-WahabEAE21}. %
NCC can successfully verify the watermarks in GPS trajectories, as demonstrated by our experiments.  
NCC of two watermarks, $w$ and $w'$, is computed as:
\begin{gather} 
    \label{eq:watermark2} %\gets
    \operatorname{NCC}(w,w') = \frac{\sum_{i}w_{i}   w'_{i}}{\sqrt{\sum_{i}w_{i}^2}\sqrt{\sum_{i}{w'_{i}}^2}}. 
\end{gather}
The value of NCC lies between $-1$ and $1$. NCC value $1$ indicates that two vectors are highly correlated,
whereas $0$ and $-1$ indicate no correlation and negative correlation, respectively. %
Finally, an average NCC score for all sub-trajectories of a given trajectory is calculated, and the verification is successful if this value is higher than the acceptance threshold $\tau$. We adopt $\tau>0.85$ based on \cite{pan2019trajguard}.
\section{Threat Model: Attacks on Trajectories}
\label{sec:attacks}

Digital watermarking is subject to adversarial attacks. 
The available knowledge limits the adversary's ability to prevent watermark verification. 
This paper assumes that an adversary has limited access, namely, knows the watermarked trajectory and the watermarking algorithm. In contrast, the original GPS data and the specific watermark embedded into the data remain unknown. 
An adversary with limited knowledge cannot remove the watermark directly. Instead, the adversary can attempt heuristic trajectory modifications to prevent watermark verification. 
We refer to such modifications as attacks on trajectories.

To quantify the utility of the trajectory modified in the adversarial settings for real-world applications, we follow the same principle as we introduced for the trajectory watermarking and apply a modification threshold $\sigma$:
\begin{align*}
\hat{\Tilde T}=AT(\Tilde{T}, \theta), \quad s.t.\ D(\Tilde{T}, \hat{\Tilde T})\leq \sigma.
\end{align*}
 Here, $AT(\cdot)$ is the attack function,
  $\Tilde{T}$ is the watermarked trajectory, 
 $\theta$ represents the specific attack parameter, $D(\cdot)$ is the distance metric, 
 $\hat{\Tilde T}$ is the modified watermarked trajectory, and $\sigma$ is the modification threshold
limiting the effects of the possible attacks on trajectories. 

In this paper, we focus on the attacks discussed in the literature in the contexts of trajectory watermarking~\citep{pan2019trajguard}, trajectory similarity measures~\citep{su2020survey} and the more general perspective of cryptography~\citep{halder2010watermarking}.
In particular, we consider four different attack types: 
noise additive attacks, point replacement attacks, size modification attacks, and the combination of these types, the hybrid attack.

\mpara{Noise Additive Attacks}
In noise additive attacks, noise is inserted into trajectory coordinates. 
\begin{enumerate}
\item \textbf{Additive Gaussian White Noise (AGWN)} 
In this attack, for each position in the trajectory, a random sample from a normal distribution is drawn and added to the GPS position.

\item \textbf{Additive Signal to Noise Ratio (ASNR)}  
This attack is similar to the previous attack, but we scale the noise to achieve a selected signal-to-noise ratio (SNR).

\item \textbf{Additive Outliers with SNR (AOSNR)} 
We randomly select points with the probability $\theta =(p_{\text{AOSNR}})$, and then add scaled noise to these positions. 

\item \textbf{Double Embedding Attack (DEA)}
In the double embedding attack, an adversary attempts to remove the original watermark by embedding a different watermark with the same approach as the original watermark. 
\end{enumerate}

\mpara{Point Replacement Attacks}
Point replacement attacks remove specific trajectory elements and replace them with information based on the adjacent points. 
\begin{enumerate}
\sloppy
\item \textbf{Replace Random Points (RRP)}
Points are selected with the probability $\theta=(p_{\text{RRP}})$,
and then those selected points are replaced with their respective previous points.
\item \textbf{Replace Random Points with Path (RRPP)}
replaces each point with the probability $\theta=(p_{\text{RRPP}})$. The replaced value is a convex combination of the remaining adjacent points.

\item \textbf{Replace Non-Skeleton Points with Path (RNSPP)}
In this attack, we use the Ramer–Douglas–Peucker (RDP) algorithm.
The points removed by the RDP algorithm are replaced with a convex combination of the adjacent points.
\end{enumerate}

\mpara{Size Modification Attacks}
In size modification attacks, the trajectory size is modified either by cropping or interpolation.
\begin{enumerate}
\item \textbf{Linear Interpolation Attack (LIA)} Additional points are inserted at random positions in the trajectory by linear interpolation, increasing the trajectory size. 
\item \textbf{Cropping Attack (CA)} Cropping attack removes selected points from the trajectory, decreasing the trajectory size.
\end{enumerate}

\mpara{Hybrid Attacks}
An adversary can combine several attacks on the same trajectory. We exemplify a hybrid attack as a sequence of a cropping attack (CA) followed by additive Gaussian white noise (AGWN) and replace random points (RRP). 

\section{Evaluation}
\label{sec:evaluation}

We aim to evaluate the effectiveness and robustness of \approach{} regarding the threat model.
In this section, we describe the experimental setup and results.

\begin{table*}[t]
  \centering
  \caption{Recognition Rate of \approach{} and baseline methods on the German and Porto datasets.}
  \begin{tabular}{cc*{13}r}
    \toprule
     \multicolumn{1}{c}{}  &\multicolumn{1}{c}{}  &\multicolumn{4}{c}{Noise additive}  & \multicolumn{3}{c}{Point replacement} & \multicolumn{2}{c}{Size mod.} &\multicolumn{1}{c}{Hybrid} & \multirow{2}{*}{Avg.} \\ 
     \cmidrule(lr){3-6} \cmidrule(lr){7-9} \cmidrule(lr){10-11}
   Method&Dataset&AGWN & ASNR &  AOSNR &DEA  & RRP  & RNSPP & RRPP & LIA& CA  &  & \\
             \midrule                                      
      SVD &German&100.0 &79.4 & 99.7&0.0 &100.0  &65.3&100.0 & 100.0&100.0&100.0 &84.3 \\
    \cline{2-13}
    (Blind) & Porto&100.0&98.2&99.3&0.0&100.0&94.7&100.0 & 100.0&100.0&100.0& 89.2 \\
   \midrule  
   IMF&German&72.5 &70.6 &74.5 &75.2 &75.8 &76.0&75.1&76.0&77.1&72.1 &74.5  \\
    \cline{2-13}
    (Non-blind)& Porto& 87.2&87.0 &90.3 &90.8 &90.1 &90.8 &90.7 & 90.3 &91.0&87.1 & 89.5\\
   \midrule  

    TrajGuard &German&87.6 &83.2 &94.4&94.4 &95.6  &74.2 &95.9 &75.2&91.9&83.8 &87.6\\
   \cline{2-13}
   (Blind) & Porto&59.8 &56.2 &55.7 &61.7 &68.3 &65.0 &68.3 &63.6  &64.5&57.5 & 62.1\\
   \midrule 
  \approach{} &German &100.0 &99.8 & 98.2&100.0 & 98.6 & 100.0& 100.0&100.0 &100.0&94.0 & 99.0\\
   \cline{2-13}
   (Non-blind)& Porto&100.0&100.0 & 99.0&100.0 & 100.0&100.0 & 100.0& 100.0 &100.0&99.9 & 99.8\\  
    \bottomrule
  \end{tabular}  
  \label{tab:example}
\end{table*}

\mpara{Datasets}
We use two real-world trajectory datasets for evaluating the proposed watermarking method. 
We randomly selected $1100$ trajectories of size $256$ from each dataset.
\begin{enumerate}
\sloppy
\item \textbf{German Dataset} is provided by a proprietary data provider. 
The dataset contains trajectory data of vehicles from two German federal states: Saxony and Lower Saxony, in September 2019. 
The average sampling rate is 12 times per minute. 
\item \textbf{Porto Dataset} 
contains variable size trajectories generated by 442 taxis from July 1, 2013, to June 30, 2014, in Porto, Portugal \citep{porto}.
The sampling rate is four times per minute. 
\end{enumerate}

\mpara{Baselines}
We adopt state-of-the-art watermarking methods from the audio domain and GPS trajectories domain.

\begin{enumerate}
\sloppy
\item \textbf{IMF Watermarking \citep{DBLP:journals/ijst/El-WahabEAE21}} is a non-blind technique used in watermarking audio signals.
Each trajectory is represented as a signal (latitude/longitude vs. time) and decomposed into multiple parts using Empirical Mode Decomposition (EMD).

\item \textbf{TrajGuard \citep{pan2019trajguard}} watermarks a GPS trajectory using a geometric transformation based on a blind scheme, i.e., it does not require the original data for the extraction.
TrajGuard partitions the trajectory into multiple parts and then distributes the 
watermark into all the sub-trajectories.

\item \textbf{SVD Watermarking ~\citep{bhat2011new}} is based on a blind audio watermarking scheme. 
This method uses Singular Value Decomposition (SVD) and quantization index modulation. 
\end{enumerate}

\mpara{Evaluation Metrics}
To assess the watermark verification effectiveness and robustness, i.e., the ability to correctly recognize a watermark in modified trajectory data, we adopt \textbf{recognition rate}. 
Recognition rate is the ratio of the number of correctly identified watermarked trajectories (true positives, $TP$) to the total number of watermarked trajectories:
$~Recognition~rate=TP/(TP+FN)$, where $FN$ is the number of false negatives, i.e., unrecognized watermarked trajectories. 
Following \cite{pan2019trajguard}, we accept the watermark to be successfully verified if the average watermark correlation between the noised trajectory and watermarked trajectory is higher than the acceptance threshold, i.e., $\tau > 85\%$.

\mpara{Evaluation Results}
\approach{} approach is effective and robust against all the considered attacks in both datasets, as shown in Table~{\ref{tab:example}}.
The average recognition rate of \approach{} is around 99\% in both datasets, confirming the effectiveness, robustness, and generalizability of \approach{}. 
Baseline methods demonstrate varying performance against some attacks across the two datasets.
For example, TrajGuard does not perform well in multiple attacks, especially on the Porto dataset. 
This is because the Porto dataset is spatially denser than the German dataset, 
making TrajGuard more vulnerable to attacks \citep{pan2019trajguard}.
Furthermore, TrajGuard embeds a smaller amount of watermark information, leading to a lower recognition rate. 
IMF watermarking failed to detect the watermark in the German dataset, whereas this method works well for the Porto dataset.
The German dataset covers a large geographical area, including two German federal states, whereas the Porto dataset is limited to one city.
A denser spatial area of the Porto dataset leads to a better decomposition and makes the verification process more effective.
Regarding the SVD watermarking, we observe that the DEA attack destroys the quantization-based watermark detection process. 
In summary, in contrast to the baselines, \approach{} is more robust against the considered attacks and less dependent on data sparsity.  

\section*{Acknowledgments} This work is partially funded by the Federal Ministry for Economic Affairs and Climate Action (BMWK), Germany, under  ``CampaNeo'' (01MD19007B), and ``d-E-mand'' (01ME19009B), the European Commission (EU H2020) under ``smashHit'' (871477), the German Research Foundation under ``WorldKG'' (424985896), and by the B-IT foundation and the state of North Rhine-Westphalia (Germany).

\balance
 
\bibliographystyle{ACM-Reference-Format}
\bibliography{ref}
\end{document}